\documentstyle[preprint,aps]{revtex}     

\begin{document}



\title{Properties of Pb(Zr,Ti)O$_3$ ultrathin films \\ 
under stress-free and open-circuit electrical boundary conditions}

\author{Emad Almahmoud, Yulia Navtsenya, Igor Kornev, Huaxiang Fu and L. Bellaiche}

\address{Physics Department,
                University of Arkansas, Fayetteville, Arkansas 72701, USA}

\date{\today}

\maketitle

\begin{abstract}
A first-principles-based scheme is developed to simulate properties
of (001) PbO-terminated Pb(Zr$_{1-x}$Ti$_{x}$)O$_3$ thin films that 
are under stress-free and open-circuit boundary conditions.
Their low-temperature spontaneous polarization never vanishes down to
 the minimal thickness, and
continuously rotates between the {\it in-plane} $<010>$ and $<110>$ directions 
when varying the Ti composition around
$x=0.50$. Such rotation dramatically enhances piezoelectricity and dielectricity. 
Furthermore, the {\it order} 
of some phase transitions changes when going from bulk to thin films.
\end{abstract}

\pacs{77.55.+f,77.80.Bh,77.65.Bn,77.22.Ch,61.50.Ah}


\narrowtext

\marginparwidth 2.7in
\marginparsep 0.5in

Ferroelectric thin films are currently of enormous technological 
interest, mostly because of 
the need in devices' miniaturization \cite{Scott}.
Many fundamental questions are still unanswered/unsettled in 
these low-dimensional systems. 
For instance, whether or not there is a critical thickness below 
which 
no ferroelectricity can occur is still under debate 
\cite{Ghosez1,Javier,Ahn,Tinte,Huaxiang2003}.
Similarly, the precise effects of surface on  
properties of thin films are opened for discussion 
\cite{Tinte,Meyer,Meyer2,Cohen2,Resta}.
One may also wonder how striking features exhibited by some bulk 
materials may evolve 
in the corresponding thin films. Typical examples of such features are 
the unusual low-symmetry phases - that are associated with a composition-induced rotation 
of the spontaneous polarization and with an enhancement
of dielectric and piezoelectric responses -- recently discovered
in the morphotropic phase boundary (MPB) of various alloys
\cite{Noheda1,PRL5427,Huaxiang2000,Rappe,Krakauer}.

One reason behind this lack of knowledge is that thin (and, particularly, 
ultrathin) films are 
difficult to synthesize in a good quality form, and their characterization
 is by no means straightforward. 
Similarly, realistically simulating thin films is a theoretical challenge.
For instance, while phenomenological  and {\it ab-initio}-based models 
have already provided a deep insight into 
thin films (see Refs. \cite{Ghosez1,PZTepol,Karin,Li} and references therein), 
such models do 
not usually incorporate some subtle surface-related phenomena -- e.g.,
charge transfer and modification of ferroelectric interactions near the surface.
On the other hand, direct first-principles techniques  can easily include such effects
 \cite{Javier,Meyer,Meyer2,Cohen2,Resta}. 
However, their large computational cost currently prevents them from being used 
to study complex phenomena 
and/or complex systems (e.g., thin films made of disordered solid solutions), especially 
at finite-temperature.
The atomistic approach of Ref.~\cite{Tinte} is a promising technique for 
investigating thin films at 
finite temperature, but its level of accuracy depends on the surface 
termination ~\cite{Tinte} -- which emphasizes that mimicking well surface 
effects on physical properties is 
tricky.

The aims of this article are twofold. First, to present a first-principles-derived 
approach allowing accurate predictions of finite-temperature properties of
 complex ferroelectric thin films -- under stress-free and open-circuit boundary conditions.
Second, to use this approach to better understand thin films by, e.g., providing 
answers to the questions mentioned above.

Here, we extend the {\it ab-initio} effective Hamiltonian scheme proposed in
Ref.~\cite{PRL5427} to mimic  thin films that (1) are made of Pb(Zr,Ti)O$_3$ (PZT); (2) 
are grown on a substrate along the [001] direction; and (3) have
{\it vacuum} above them.
More precisely, the total energy $E_{tot}$ of such low-dimensional systems is written as:
\begin{eqnarray}
   E_{tot} ( \{ { \bf u}(i) \},\{ { \bf v}(i) \}, \eta,
     \{ \sigma_{\it i} \})   =
   E_{mat} (  \{ { \bf u}(i) \},\{ { \bf v}(i) \}, \eta,\{ \sigma_{\it i} \}) \nonumber \\
    \quad +~~P\sum_{j} u_{z}(j)~+~T\sum_{j} v_z(j)~+~S\sum_{j} \sum_{\alpha=x,y} u_{\alpha}(j) 
    (u_{\alpha}(j+\hat{\alpha}) +u_{\alpha}(j-\hat{\alpha})), 
   \end{eqnarray}
where ${\bf u}(i)$ are the (B-centered) local soft modes in unit cells $i$ of the film, and are 
directly proportional to the electrical polarization. 
${\bf v}(i)$ are inhomogeneous strains around the $i$ site, while
$\eta$ is the homogeneous strain tensor. $\{ \sigma_{\it i} \}$
characterizes the alloy configuration \cite{PRL5427}.
$E_{mat}$ represents the intrinsic ferroelectric and elastic interactions {\it inside} the film, with
its analytical expression and first-principles-derived parameters being those of PZT {\it bulks} 
\cite{PRL5427}. Only four (out of 26) parameters are composition-dependent 
in $E_{mat}$: they are those associated with the so-called local-mode self energy \cite{PRL5427}.
The last three terms  of Eq~(1) 
mimic explicit interactions between this film and the vacuum, with  
the $j$ index running over all the B-sites that are the closest to the free surface.
$u_{x}(j)$, $u_{y}(j)$ and $u_{z}(j)$ denote 
the (x-, y- and z-) Cartesian component of ${\bf u}(j)$ along the pseudo-cubic [100], 
[010] and [001] directions, respectively.
$\alpha$ runs over the x- and y-axes (i.e., it does not include the growth direction).
$u_{\alpha}(j+\hat{\alpha})$ (respectively, $u_{\alpha}(j-\hat{\alpha})$) is 
the $\alpha$-component of the local mode 
centered on the B-site that is the closest from the $j$ site that is parallel 
(respectively, antiparallel) to the $\alpha$-axis.
The $P$  and $T$  parameters quantify how vacuum affects the {\it out-of-plane} 
components ($u_z$ and $v_z$) 
of the local modes and inhomogeneous strains near the surface, respectively. 
$S$ characterizes the change, with respect to the bulk, of 
the short-range interaction between the {\it in-plane} components of the local 
modes near the surface.
The $P$, $T$ and $S$ parameters are determined from 
first-principles calculations on a {\it  PbO-terminated} (001) 17-atom slab
 (corresponding to 3 B-O and 4 A-O atomic layers) 
of a PZT alloy, as mimicked by the virtual crystal alloy approximation \cite{PRB7877}, 
surrounded by a  vacuum region being 
2 lattice-constants thick \cite{footnoteref3}. Using Eq~(1)  with these
parameters 
results in an excellent agreement with first-principles predictions for the 
layer-by-layer profile of the polarization in various
PbO-terminated slabs (e.g., in slabs that are {\it not} used in the fitting 
procedure) \cite{footnote1}.

Furthermore, the substrate is assumed to be {\it inert}, i.e. no term 
analogous to the last three expressions
of Eq.~(1) is considered at the substrate/film interface. 
The local modes and inhomogeneous strain-related variables are also 
forced to vanish in the substrate (as well as, 
of course, in the vacuum), being consistent with the fact that some 
commonly-used substrates (e.g., MgO) are ferroelectric inactive.
As a result, dipole-dipole interactions only occur in the film, 
which automatically guarantees the generation of
a depolarizing field inside the film if this latter has 
a component of its electrical polarization along the 
growth direction. Our films are thus under {\it open-circuit 
electrical boundary conditions}.
Technically, we use the total energy of Eq.~(1) in Monte-Carlo simulations 
to compute the finite-temperature properties of PbO-terminated PZT ultrathin
 films having a number of layers -- to be denoted by $d$ -- 
ranging between 1 and 6. These simulations use the Metropolis algorithm \cite{metropolis},
and provide the thermodynamically equilibrated local soft modes and strains as outputs for each temperature.
The strains (like the local  modes, but unlike the alloy configuration that is kept frozen)
are allowed to fully relax. Our studied films are thus {\it stress-free} 
in terms of mechanical boundary conditions. 
Typically, we use a huge number of Monte-Carlo sweeps (up to 20 millions) 
and large $10\times 10\times 40$ periodic supercells --- 
the regions made of substrate and vacuum are thus altogether $40-d$ lattice 
constant thick along
the growth direction --- to get well-converged results. The B-atoms are 
randomly distributed in the film.

Figure~1 displays the predicted Cartesian
coordinates ($<u_{x}>$, $<u_{y}>$ and $<u_{z}>$) of the average of the local 
mode vectors in the $\simeq$ 20\AA-thick ($d$=5) PZT film having  a 
50\% Ti composition, 
as a function of the rescaled temperature \cite{footnote2}. Two sets of
 calculations are reported: 
one incorporating all the surface-related terms of Eq.~(1) {\it vs.} 
another for which the $P$, $T$ and $S$ coefficients
are all turned off. In both cases, each coordinate is close to zero 
at high temperature, indicating
a paraelectric phase. As the temperature is cooled down and passes 
through a T$_c$ critical temperature,  
$<u_{y}>$ jumps -- and then increases when the temperature further
 decreases --, while $<u_{x}>$ 
and $<u_{z}>$ remain nearly null \cite{footnote3}. This characterizes
 a transition to a ferroelectric phase having a 
polarization pointing along a $<100>$ direction, as in the PZT bulk 
having the same Ti composition of 50\%  \cite{PRL5427}.
However, the polarization direction in the
 film can be along either the $y$-axis (as in Fig.~1)
or the $x$-axis, but (unlike in the bulk) {\it never} along the [001] growth direction 
($z$-axis) -- in order
to avoid the generation of a huge depolarizing field \cite{Tinte}. 
Interestingly, we also found (not shown here) that thinner PbO-terminated 
films, down to $d$=1, also exhibit
a non-zero in-plane polarization in their ground-state. Our 
calculations thus predict 
that {\it stress-free PZT films, under open-circuit 
electrical boundary conditions
and with Ti compositions around 50\%,
do {\it not} have any critical thickness below which 
ferroelectricity disappears}! Such conclusion seems to contradict 
recent studies (see, e.g. Ref~\cite{Javier}).
However, one has to realize that, e.g., Ref.~3 defines the critical thickness with respect to the vanishing
of {\it the component of the polarization along the z-axis} -- while we define it here with respect to the 
vanishing of {\it all} the components of the polarization --
and studies thin films that are under a compressive strain
and  close (but not equal to) short-circuit electrical 
boundary conditions.
The specificity of our present study
does not allow us to generalize our findings for the mechanical  
and electrical boundary conditions associated with some of 
these previous studies.

Figure 1 further tells us that the T$_c$ Curie temperature of
 the $d=5$ PZT film is  around $525 \,K$, that is
lower by $\simeq$ 120 $\,K$ from the one of the corresponding
 PZT bulk \cite{Jpn196}, when turning on
$P$, $T$ and $S$ in Eq~(1). 
Neglecting such parameters leads to a much smaller T$_c$($\simeq 370\,K$).
Such large difference in T$_c$ is found to be 
mostly due to the $S$ parameter (which characterizes 
the vacuum-induced change of short-range interaction near the surface).
 Interestingly, this
 $S$ parameter is rather sensitive to the surface termination~\cite{Meyer}. 
This implies that 
properties of {\it ultrathin films} having two differently-terminated surfaces can 
dramatically differ, 
even if those films are made from the same material and have similar thickness.

Figure 2 sheds more light on surface effects by displaying the layer-by-layer
 profiles of the 
local modes for the $d=5$ ultrathin film having a Ti composition of 50\%, when 
turning on and off 
the $P$, $T$ and $S$ parameters at 10K.
One can clearly see that, in both cases, the $y$-component of the local
 dipoles centered 
in the B-layers, that are neither located near the interface nor the surface, 
has a more-or-less layer-independent value that is slightly smaller than the
 one in the bulk ($\simeq$  0.107 a.u.).
On the other hand, the B-layer that is the closest to the substrate/film 
interface has much
smaller in-plane dipoles. The $P$, $S$ and $T$ 
parameters also have a significant effect
on the $y$-component  of the dipole in the B-layer located near the
 vacuum: neglecting these parameters
generates a small value that is comparable to the one near the 
interface, while turning them on
yields a large enhancement of this $y$-component in this 
PbO-terminated ultrathin film. Such enhancement is due to the $S$ coefficient, and
 explains why the Curie temperature of the film increases 
 when turning on
the surface-related parameters -- since it is commonly accepted
 that the larger the spontaneous
polarization is at small temperature (see Fig~2) the higher
 T$_c$ (see Fig~1) \cite{phi4}. 
Another difference worthy to be noticed between the two kinds 
of calculations is about
the $z$-component of the dipole in the B-layer located near 
the vacuum: the fitted value of the $P$ parameter leads to
a $u_z$ that does not vanish anymore and that is negative 
-- mostly characterizing Pb-displacements oriented {\it inwards} the film near
the PbO-terminated surface, as consistent with Ref.~\cite{Meyer2}.

We now use our scheme to investigate properties of 
Pb(Zr$_{1-x}$Ti$_{x}$)O$_{3}$ films, 
with a fixed thickness ($d$=5) at 10K, {\it as a function of 
Ti composition};
all the surface relaxations of Eq~(1) being included. 
Figure 3a shows that the spontaneous polarization is nearly 
parallel to a $<010>$ direction, that is
 perpendicular to the growth direction,
for Ti compositions larger than 49.6\% \cite{footnote3}. The 
associated phase, to be denoted by $T$, can be considered
to be tetragonal when solely focusing on the film (otherwise, 
it is orthorhombic 
when including 
the substrate and vacuum in addition to the film). This polarization 
becomes nearly parallel to a $<110>$ pseudo-cubic
direction \cite{footnote3} -- i.e., it changes of {\it in-plane} 
direction  -- for $x$ smaller than 47.0\%. This yields
a phase that can be classified as orthorhombic and will be denoted 
as $O$. 
One striking feature of Fig.~3a is the behavior of the local mode for 
compositions decreasing between 49.6\% and 47.0\%: 
$<u_{y}>$ decreases, while $<u_{x}>$ smoothly increases and $<u_{z}>$
 remains nearly null
\cite{footnote3}. 
This characterizes a continuous (in-plane) rotation of the polarization 
within a (001) plane from a $<010>$
 to a $<110>$ direction, 
with a resulting phase adopting a monoclinic M$_C$ symmetry \cite{DavidMorrel}!
Interestingly, the phase diagram depicted in Fig.~3a can be rather well {\it 
qualitatively and quantitatively} 
deduced from the one occurring in the MPB of PZT {\it bulks}, when imposing 
the additional constraint that $<u_{z}>$=0 to these latters. As a matter of fact,
 at small-temperature, the polarization of PZT bulks 
rotates from a  $<010>$ (tetragonal-associated) to the  $<111>$ 
(rhombohedral-associated) direction,
via a newly-discovered monoclinic M$_A$ phase (characterized by a 
polarization lying along 
 $<uvu>$ with $u < v$ \cite{Noheda1}),
for Ti compositions decreasing between 49.5\%  and 47.0\%  \cite{PRL5427}.
One noticeable difference exists, however, for the {\it nature} of the 
composition-induced phase transition around 47.0\% 
between the bulks and the films:
the rhombohedral R--to-- monoclinic M$_A$ phase transition occurring in
 the bulk is of first-order \cite{PRL5427,DavidMorrel}, while
the orthorhombic O--to--monoclinic M$_C$ transition appears to be of
 second-order in the film (see the continuous 
rotation of the mode in Fig.~3a). Interestingly, such difference is
consistent with the
sixth-order Devonshire theory for the R--to--M$_A$  and O--to--M$_C$ 
transitions \cite{DavidMorrel}.

Fig.~3b and ~3c reveal that -- when representing the dielectric and piezoelectric tensors in the basis 
formed by the x-, y- and z-axes -- the d$_{16}$ 
piezoelectric  and $\chi_{11}$ dielectric coefficients  of PZT films not only peak at both transitions 
(as consistent with phase-transition theory) but also adopt 
remarkably large values anywhere between 47.0\% and 49.5\%.
As in bulks \cite{PRL5427,Huaxiang2000,Krakauer}, these large values
 result from the composition-induced rotation of the polarization.
PZT ultrathin films may thus be promising to design miniaturized 
ferroelectric devices with large  dielectric and piezoelectric responses.

In summary, we developed a first-principles-based scheme to investigate
finite-temperature properties
of (001) PZT stress-free ultrathin films under open-circuit electrical
 boundary conditions, that are PbO-terminated
and have a Ti composition around 50\%. 
We found that (1) such films do {\it not} have any critical thickness 
below which ferroelectricity disappears; (2) their polarization are 
perpendicular to the growth
direction; (3) surface effects, and especially vacuum-induced changes 
of short-range interaction,  significantly affect
their local and macroscopic properties; (4) 
these ultrathin films exhibit a morphotropic phase boundary where their  
polarization continuously rotates,
in a (001) plane between a $<010>$ and $<110>$ direction, as the Ti 
composition decreases; (5) such rotation leads
to large piezoelectricity and dielectricity; and (6) the nature of 
phase transitions can change when going from bulks to
ultrathin films.

We thank I. Naumov for useful discussions.
This work is supported by ONR grants N 00014-01-1-0365, N00014-04-1-0413
 and 00014-01-1-0600, and
by NSF grants DMR-0404335 and DMR-9983678.

\narrowtext

\begin{figure}
\caption{Film average of the cartesian coordinates 
($<u_{x}>$, $<u_{y}>$ and $<u_{z}>$) 
of the local modes in the $\simeq$ 20\AA-thick  ($d$=5)
 PZT film having a 50\% Ti composition, as a function of
the rescaled temperature. 
The filled symbols represent predictions when all the 
surface-related parameters of Eq.~(1) are turned on, while
the open symbols show the corresponding results when 
neglecting the $P$, $T$ and $S$ parameters.
The y-component of the local modes is indicated via circles, while
the triangles and squares display the x- and z-components, 
respectively, of these local modes. 
(The z-component 
of the local modes is 
always null for each temperature and is not shown for clarity, when 
neglecting the $P$, $T$ and $S$ parameters.)}
\end{figure}

\begin{figure}
\caption{Local modes for the $d=5$ PZT ultrathin film as a 
function of the layer index along the growth direction, at 10K.
An index of $1$ characterizes the B-layer that is closest 
to the substrate, while the B-layer located near
the vacuum has an index of $5$. 
The solid line and filled symbols represent predictions when
 all the surface-related parameters of Eq.~(1) are turned on, while
the dashed line and open symbols show the corresponding 
results when neglecting the $P$, $T$ and $S$ parameters.
The y-component of the local modes is indicated via circles, while
the squares display the z-component of these local modes. 
(The x-component of the local modes is always null in each
layer, and is not shown for clarity.)
The simulated film has a composition of 50\% of Ti.}

\end{figure}

\begin{figure}
\caption{Properties of the the $d=5$ ultrathin PZT films as a 
function of Ti composition, at 10K.
Panel (a) displays the film average of the cartesian coordinates 
($<u_{x}>$, $<u_{y}>$ and $<u_{z}>$) 
of the local modes. Panel (b) shows the d$_{22}$ and d$_{16}$ 
piezoelectric
coefficients. Panel (c) displays the $\chi_{11}$ and $\chi_{22}$ 
dielectric susceptibilities.
All the parameters of Eq.~(1) are turned on in these simulations.
All the calculations are performed using $10\times 10\times 40$ supercells, 
except those  corresponding to a Ti composition around 0.47 -- 
for which $20\times 20\times 40$ supercells are used to confirm that some 
piezoelectric and dielectric coefficients 
peak around this composition, as expected by phase-transition theory.}
\end{figure}


\begin{references}
\bibitem{footnoteref3} We obtained $P$=+0.0144, $T$=+0.00693
and $S$=-0.00405 in atomic units. These values were numerically 
found to be independent on the lateral size, as well as
on the number of vacuum layers, of the supercells used within the approach of Eq~(1) to 
fit the first-principles calculations on the 17-atom slab. 
We also assume that these parameters are composition-independent since none of them
is associated with the local-mode self-energy (which is the energy that is
rather sensitive to concentration in PZT bulks \cite{PRL5427}).
\bibitem{footnote1} Note that Ref.\cite{Meyer2} 
predicts that the PbO-terminated surface
is the most stable one in PbTiO$_3$-like compounds.
\bibitem{footnote3} Note that, in fact, $<u_{z}>$ is slightly negative,
when all the parameters of Eq.~(1) are turned on. For 
the sake of simplicity, we 
neglect such small value in the discussion.
\end{references}
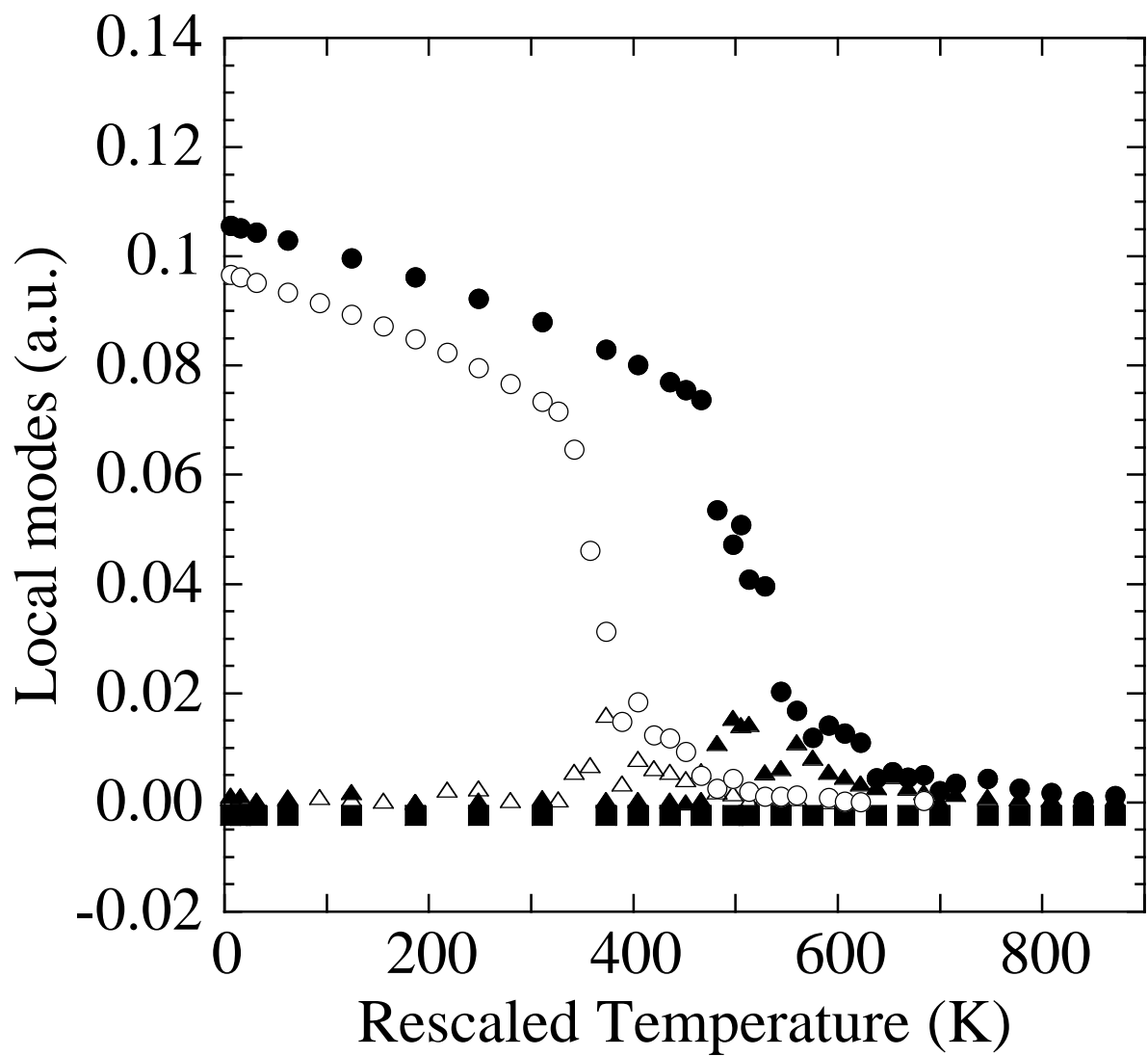
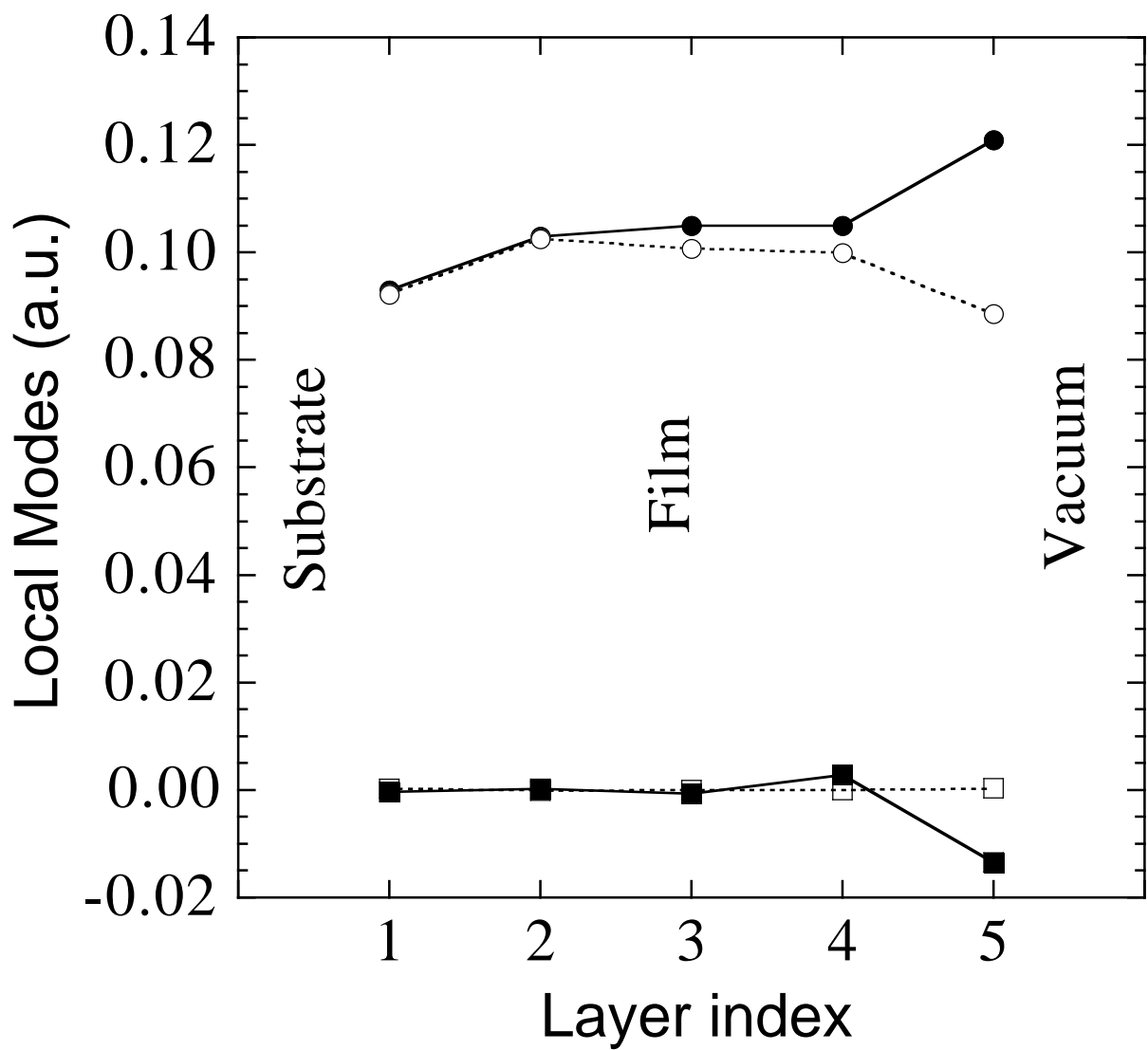
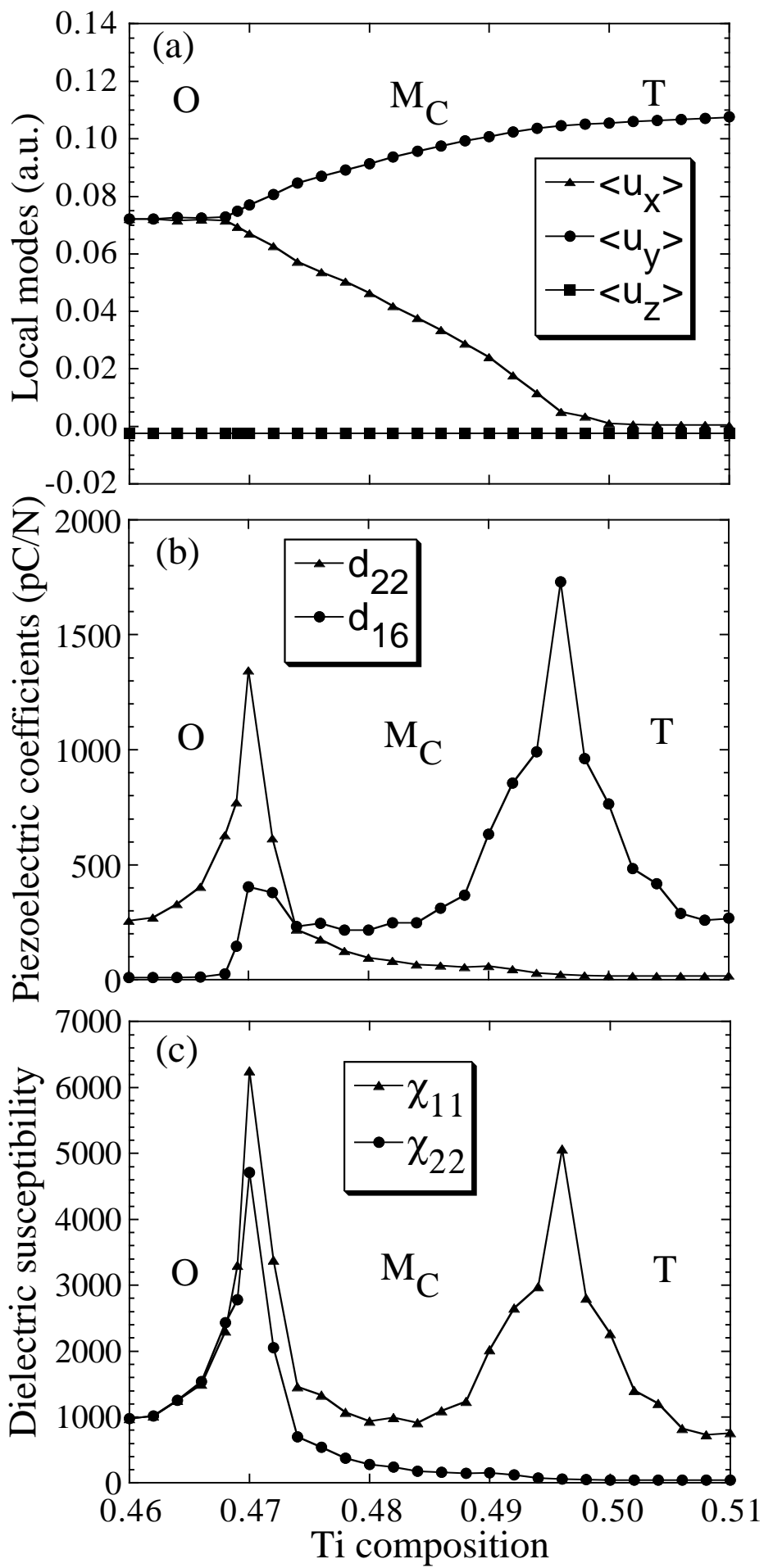
\end{document}